\documentclass[apj]{emulateapj}
\usepackage{mathptmx}

\newcommand{\ltsim}{\raisebox{-.5ex}{$\;\stackrel{<}{\sim}\;$}}
\newcommand{\gtsim}{\raisebox{-.5ex}{$\;\stackrel{>}{\sim}\;$}}
\newcommand{\kms}{\ifmmode {\rm km\ s}^{-1} \else km s$^{-1}$\fi}
\newcommand{\et}{et al.\ }
\newcommand{\aox}{$\alpha_{\rm ox}$}
\newcommand{\nh}{$N_{\rm H}$}
\newcommand{\Ka}{Fe K$\alpha$}
\newcommand{\xmm}{{\hbox{\sl XMM-Newton}}}
\newcommand{\chandra}{{\sl Chandra}}

\journalinfo{The Astrophysical Journal, 
630:??--??, 2005 September 10, astro-ph/0505482}
\slugcomment{Received 2005 April 8; accepted 2005 May 20}

\shorttitle{X-RAY PROPERTIES OF LUMINOUS HIGH-Z AGNS}
\shortauthors{SHEMMER ET AL.}

\begin{document}
\title{THE X-RAY SPECTRAL PROPERTIES AND VARIABILITY OF LUMINOUS HIGH-REDSHIFT \\
ACTIVE GALACTIC NUCLEI}
\author{
O.~Shemmer, \altaffilmark{1}
W.~N.~Brandt,\altaffilmark{1}
C.~Vignali,\altaffilmark{2,3}
D.~P.~Schneider,\altaffilmark{1} \\
X.~Fan,\altaffilmark{4}
G.~T.~Richards,\altaffilmark{5} and
Michael~A.~Strauss\altaffilmark{5}
}

\altaffiltext{1}
                {Department of Astronomy \& Astrophysics, The Pennsylvania
                State University, University Park, PA 16802, USA;
                ohad@astro.psu.edu}

\altaffiltext{2}
                {INAF - Osservatorio Astronomico di Bologna, Via Ranzani 1,
                40127 Bologna, Italy}

\altaffiltext{3}
                {Dipartimento di Astronomia, Universita` degli Studi di
                Bologna, Via Ranzani 1, 40127 Bologna, Italy}

\altaffiltext{4}
               {Steward Observatory, University of Arizona, 933 North Cherry
               Avenue, Tucson, AZ 85721, USA}

\altaffiltext{5}
                {Princeton University Observatory, Peyton Hall, Princeton,
                NJ 08544, USA}

\begin{abstract}
We perform a detailed investigation of moderate-to-high quality \hbox{X-ray} spectra
of ten of the most luminous active galactic nuclei (AGNs) known at \hbox{$z>4$} (up to
\hbox{$z\sim6.28$}). This study includes five new \xmm\ observations and five archived
\hbox{X-ray} observations (four by \xmm\ and one by \chandra).
We find that the \hbox{X-ray} power-law photon indices of our sample,
composed of eight radio-quiet sources and two that are moderately radio loud,
are not significantly different from those of lower redshift AGNs.
The upper limits obtained on intrinsic neutral hydrogen column densities,
\hbox{\nh$\ltsim$10$^{22}$--10$^{23}$~cm$^{-2}$}, indicate that these AGNs are not
significantly absorbed. A joint fit performed on
our eight radio-quiet sources, with a total of $\sim$7000 photons,
constrains the mean photon index of \hbox{$z>4$} radio-quiet AGNs to
$\Gamma$=1.97$^{+0.06}_{-0.04}$, with no detectable intrinsic
dispersion from source to source. We also obtain a strong constraint on the mean
intrinsic column density, \nh$\ltsim$3$\times$10$^{21}$~cm$^{-2}$,
showing that optically selected radio-quiet AGNs at \hbox{$z>4$} are, on average, not
more absorbed than their lower-redshift counterparts. All this suggests that the X-ray
production mechanism and the central environment in radio-quiet AGNs have not
significantly evolved over cosmic time. The mean equivalent width of a putative
neutral narrow \Ka\ line is constrained to be $\ltsim$190~eV, and similarly we place
constraints on the mean Compton reflection component ($R\ltsim$1.2). None of the
AGNs varied on short ($\sim$1~hr) timescales, but on longer timescales
(months-to-years) strong variability is observed in four of the sources. In
particular, the \hbox{X-ray} flux of the $z$=5.41 radio-quiet AGN SDSS~0231$-$0728
dropped by a factor of $\sim$4 over a rest-frame period of 73~d. This is the most
extreme \hbox{X-ray} variation observed in a luminous \hbox{$z>4$} radio-quiet AGN.
\end{abstract}


\section{Introduction}
\label{introduction}

Active galactic nuclei (AGNs) at \hbox{$z>4$} are valuable cosmological probes of the
physical environment in the $\sim$1~Gyr old Universe, following the
re-ionization epoch at the end of the `dark ages' (e.g., Fan \et 2002).
Investigations of physical processes in these sources, and in particular their
energy production mechanism, are important for understanding
the ionizing flux contribution from AGNs in the early Universe, black hole
(BH) growth, and the evolution of mass accretion in galactic nuclei
(see Brandt \& Hasinger 2005 for a recent review).
It is not yet clear whether the small-scale physics of active nuclei are sensitive
to the strong large-scale environmental changes the Universe has experienced
since $z\approx6$.

The penetrating \hbox{X-ray} emission from AGNs is of particular interest, since it
allows one to probe the innermost region of the accretion flow, near the central
supermassive BH. Until fairly recently, \hbox{X-ray} detections of \hbox{$z>4$} AGNs
were limited to a handful of sources with poor photon statistics, preventing a detailed
analysis of their X-ray spectral properties (Kaspi, Brandt, \& Schneider 2000); this
was also, in part, a consequence of the relatively small number ($\ltsim$100) of such
sources discovered by the year 2000. The past five years have seen a rapid increase in
optical discoveries of \hbox{$z>4$} AGNs, largely as a result of two surveys, the
Palomar Digital Sky Survey (DPOSS; Djorgovski \et 1998) and the Sloan Digital Sky
Survey (SDSS; York \et 2000), which have raised the number of such sources to
$\sim$600 (e.g., Schneider \et 2005). A rapid increase in \hbox{X-ray} detections of
\hbox{$z>4$} AGNs soon followed, with $\sim$90 such sources to date.\footnote{See
http://www.astro.psu.edu/users/niel/papers/highz-xray-detected.dat
for a regularly updated compilation of \hbox{X-ray} detections at \hbox{$z>4$}.}
\chandra\ and \xmm\ observations of these sources have provided constraints on
their mean global \hbox{X-ray} spectral properties, in particular their photon index
and intrinsic absorption column density (e.g., Brandt \et 2004; Vignali \et 2005).
However, many of the \hbox{X-ray} observations are shallow, aimed at just detecting
the AGN, and do not allow useful measurements of the spectral properties
in individual sources (e.g., Brandt \et 2002; Vignali \et 2003a, 2005). Exposure
times of several tens of ks with \chandra\ and \xmm\ are needed to measure the spectral
properties accurately. Such observations are important for investigating the energy
production mechanism at this early epoch, and testing whether the \hbox{X-ray} spectral
properties have evolved over cosmic time.

\begin{deluxetable*}{lccccccccc}
\tablecolumns{10}
\tablewidth{460pt}
\tablecaption{Log of \xmm\ Observations \label{obs_log}}
\tablehead
{
\colhead{} &
\colhead{} &
\colhead{$\alpha$} &
\colhead{$\delta$} &
\colhead{Galactic \nh\tablenotemark{a}} &
\colhead{Obs.} &
\multicolumn{3}{c}{Exp. Time (ks) / Total Counts} &
\colhead{} \\
\colhead{AGN} &
\colhead{$z$} &
\colhead{(J2000.0)} &
\colhead{(J2000.0)} &
\colhead{(10$^{20}$~cm$^{-2}$)} &
\colhead{Date} &
\colhead{MOS1} &
\colhead{MOS2} &
\colhead{pn} &
\colhead{Ref.}
}
\startdata
\cutinhead{New Observations}
PSS 0121$+$0347\tablenotemark{b}  & 4.13 & 01 21 26.2 & $+$03 47 06.3 & 3.23 &
2004 Jan 09 & 25.7 / 139 & 25.9 / 185 & 21.7 / 540 & 1 \\
SDSS 0210$-$0018\tablenotemark{b} & 4.77 & 02 10 43.2 & $-$00 18 18.3 & 2.66 &
2005 Feb 05 & 18.7 / 112 & 18.7 / 117 & 16.5 / 594 & 1 \\
SDSS 0231$-$0728 & 5.41 & 02 31 37.7 & $-$07 28 54.5 & 2.90 &
2004 Jan 07 & 31.6 / 70 & 31.6 / 101  & 27.7 / 309 & 1 \\
PSS 0926$+$3055  & 4.19 & 09 26 36.3 & $+$30 55 04.9 & 1.89 &
2004 Nov 13 & 27.8 / 289 & 27.8 / 295 & 23.8 / 1156& 1 \\
PSS 1326$+$0743  & 4.17 & 13 26 11.8 & $+$07 43 57.5 & 2.01 &
2003 Dec 28 & 31.3 / 247 & 31.3 / 229 & 27.1 / 963 & 1 \\
\cutinhead{Archival Observations}
Q~0000$-$263 & 4.10 & 00 03 22.9 & $-$26 03 17 & 1.67 &
2002 Jun 25 & 34.0 / 295 & 34.0 / 322 & 34.0 / 1229 & 2 \\
BR~0351$-$1034    & 4.35 & 03 53 46.9 & $-$10 25 19 & 4.08 &
2002 Aug 23 & 17.9 / 31  & 17.9 / 15  & 15.1 / 166  & 3 \\
SDSS~1030$+$0524  & 6.28 & 10 30 27.1 & $+$05 24 55 & 3.17 &
2003 May 22 & 44.3 / 80  & 43.9 / 81  & 43.4 / 333  & 4 \\
BR~2237$-$0607    & 4.56 & 22 39 53.6 & $-$05 52 19 & 3.84 &
2003 May 17 & 17.2 / 60  & 17.3 / 76  & 16.3 / 306  & 5 \\
\enddata
\tablenotetext{a}{Galactic absorption column densities from
Dickey \& Lockman (1990), obtained using the optical coordinates of the
sources given in this Table and using the HEASARC \nh\ tool at
http://heasarc.gsfc.nasa.gov/cgi-bin/Tools/w3nh/w3nh.pl}
\tablenotetext{b}{Moderately radio-loud source.}
\tablerefs{1. This work; 2. Ferrero \& Brinkman (2003); 3. Grupe \et (2004);
4. Farrah \et (2004); 5. Grupe \et (2005)}
\end{deluxetable*}

Recent studies of \hbox{X-ray} spectral properties and their dependence on
redshift and luminosity have led to conflicting conclusions. For example, while
Bechtold \et (2003) reported that the \hbox{X-ray} photon indices ($\Gamma$)
of \hbox{$z>4$} AGNs are {\em flatter} (1.5$\pm$0.5) than those of nearby AGNs,
Grupe \et (2005) reported that their $\Gamma$ are rather {\em steep} (2.23$\pm$0.48).
The first study was based upon a sample of 16 radio-quiet sources observed by
\chandra\ with low total counts (see Table~1 of Bechtold \et 2003), and one of their
objects (SDSS~J103027.10$+$052455.0, hereafter SDSS~1030$+$0524) is common with the
present work. The second study was based upon a sample of 16 sources (11 of which are
radio quiet) observed with \xmm\ with low-to-high total counts (see Table~1 of Grupe
\et 2005); four of those sources (Q~0000$-$263, BR~0351$-$1034, SDSS~1030$+$0524,
and BR~2237$-$0607) are common with the present work, and we use the same \hbox{X-ray}
data that were discussed in that study. The recent analysis by Vignali \et (2005),
which is based upon the joint \hbox{X-ray} spectral fitting of a sample of 48
\hbox{$z>4$} radio-quiet AGNs observed with \chandra\ in the redshift range 3.99--6.28,
argues that the \hbox{X-ray} spectral properties of AGNs do not show any significant
evolution or luminosity dependence. Our aim in this paper is to address this puzzle by
uniformly analyzing a sample of moderate-to-high quality \hbox{X-ray} spectra of
\hbox{$z>4$} AGNs, with $\approx$200--1000 counts per source. In this paper, we
consider radio-quiet and radio-loud AGNs separately, as radio-loud sources might
exhibit \hbox{X-ray} emission from their jets. The \hbox{X-ray} properties of the two
AGN sub-classes differ significantly from one another; for example, the \hbox{X-ray}
spectra of radio-loud AGNs are flatter than those of radio-quiet sources (e.g., Wilkes
\& Elvis 1987; Cappi \et 1997; Porquet \et 2004) and, at high redshift,
radio-loud AGNs show signs of higher intrinsic absorption (e.g., Worsley \et 2004a, b)
and larger \hbox{X-ray} variability (e.g., Fabian \et 1999). Since radio-quiet sources
represent the majority of luminous AGNs at all redshifts (e.g., Stern \et 2000;
Ivezi{\' c} \et 2002; Petric \et 2003), our analysis focuses on the properties of
this sub-class.

Moderate-to-high quality spectra of only five \hbox{$z>4$} radio-quiet AGNs
have been obtained with \xmm\ and \chandra\ over the past three years, with
only one such source having more than 1000 counts (Q~0000$-$263, Ferrero
\& Brinkmann 2003; see also Farrah \et 2004; Grupe \et 2004; Schwartz \& Virani
2004; Grupe \et 2005). In this paper we almost double that number with new data from
\xmm, adding two more sources with over 1000 counts, and study the broad-band
X-ray spectral properties of ten AGNs with an average redshift of 4.8. In
\S~\ref{observations} we describe our five \xmm\ observations, the data reduction, and
the analysis. In \S~\ref{expanded} we expand our original sample by adding archived
high-quality \xmm\ and \chandra\ spectra of five \hbox{$z>4$} radio-quiet AGNs, and in
\S~\ref{joint} we investigate the mean \hbox{X-ray} spectral properties of our
radio-quiet sample. Our results are discussed in \S~\ref{discussion} and summarized in
\S~\ref{summary}. Throughout the~paper we use the standard cosmological
model, with parameters $\Omega_{\Lambda}$=0.7, $\Omega_{\rm M}$=0.3, and
$H_0$=70~\kms~Mpc$^{-1}$ (Spergel \et 2003).

\section{New High-Quality Spectra of \hbox{$z>4$} AGNs}
\label{observations}
\subsection{Observations and Data Reduction}
\label{reduction}

We have obtained imaging spectroscopic observations of five \hbox{$z>4$} AGNs with
\xmm\ (Jansen \et 2001). Three of the sources were discovered in the DPOSS
survey and were detected~in X-rays by Vignali \et (2003a); two were discovered
in the SDSS survey and detected in~X-rays by Vignali \et (2001, 2003b).
Three of the sources are radio quiet with an upper limit on the radio flux,
while two, PSS~0121$+$0347 and SDSS~0210$-$0018, are moderately radio loud, following
the classification of Kellermann \et (1989), and have $R$=300 and $R$=80,
respectively (Vignali \et 2003a, 2001). These five sources were among the X-ray
brightest of a sample of \hbox{$z>4$} AGNs previously targeted with \chandra\
snapshot ($\ltsim$10~ks) observations (Vignali \et 2001, 2003a, b). One of these
sources, SDSS~0231$-$0728 at $z$=5.41 (Anderson \et 2001), is the highest redshift AGN
detected with the automated SDSS detection algorithm. The short, higher
angular resolution \chandra\ observations allow prediction of the number of counts
expected from the \xmm\ observations, and they eliminate the possibility of
significant source confusion due to the lower angular resolution of \xmm. No
nearby contaminating sources were detected in the \chandra\ images of our five~AGNs.

\begin{deluxetable}{lcccc}
\tablecolumns{5}
\tablecaption{Best-Fit \hbox{X-ray} Spectral Parameters \label{best_fit}}
\tablehead
{
\colhead{AGN} &
\colhead{\nh\tablenotemark{a}} &
\colhead{$\Gamma$} &
\colhead{$f_{\rm E}$(1~keV)\tablenotemark{b}} &
\colhead{$\chi^{2}$(DOF)}
}
\startdata
PSS~0121$+$0347  & $\le2.91$  & $1.81^{+0.16}_{-0.16}$ &
$7.64^{+0.96}_{-1.01}$        & $54.8(52)$  \\ \\
SDSS~0210$-$0018 & $\le4.17$ & $1.81^{+0.15}_{-0.14}$ &
$6.45^{+0.78}_{-0.78}$        & $64.2(75)$  \\ \\
SDSS~0231$-$0728 & $\le19.90$ & $1.85^{+0.33}_{-0.31}$ &
$2.10^{+0.59}_{-0.58}$        & $25.3(43)$  \\ \\
PSS~0926$+$3055  & $\le1.02$  & $1.99^{+0.08}_{-0.08}$ &
$17.58^{+1.15}_{-1.15}$       & $76.1(78)$  \\ \\
PSS~1326$+$0743  & $\le0.47$  & $1.87^{+0.10}_{-0.10}$ &
$12.54^{+0.84}_{-1.13}$       & $51.2(65)$  \\  \\
\hline \\
Q~0000$-$263 & $\le0.40$ & $2.02^{+0.07}_{-0.07}$ &
$12.59^{+0.75}_{-0.75}$ & $71.3(82)$ \\ \\
BR~0351$-$1034    & $\le5.29$ & $1.76^{+0.31}_{-0.26}$ &
$5.24^{+0.96}_{-0.97}$  & $8.7(10)$  \\ \\
SDSS~1030$+$0524  & $\le7.77$ & $1.92^{+0.23}_{-0.21}$ &
$1.39^{+0.23}_{-0.23}$  & $40.9(44)$\\ \\
BR~2237$-$0607    & $\le3.30$ & $2.09^{+0.22}_{-0.21}$ &
$3.82^{+0.59}_{-0.57}$  & $20.4(25)$ \\
\enddata
\tablecomments{The best-fit photon index, normalization, and $\chi^2$ were
obtained from a model consisting of a Galactic-absorbed power-law without
any intrinsic absorption.}
\tablenotetext{a}{Intrinsic column density in units of 10$^{22}$~cm$^{-2}$.
Upper limits were calculated
using the intrinsically-absorbed power-law model with Galactic absorption,
and represent 90\% confidence limits for each value,
taking one parameter of interest ($\Delta \chi^{2}$=2.71; e.g., Avni 1976).}
\tablenotetext{b}{Power-law normalization for the pn data, taken from
joint fitting of all three EPIC detectors with the Galactic-absorbed
power-law model; units are 10$^{-6}$~keV~cm$^{-2}$~s$^{-1}$~keV$^{-1}$.}
\end{deluxetable}

A log of the new \xmm\ observations appears in Table~\ref{obs_log} (first five sources).
The data were processed using standard {\sc sas\footnote{\xmm\ Science Analysis System.
See http://xmm.vilspa.esa.es/ external/xmm\_sw\_cal/sas\_frame.shtml} v6.1.0} and {\sc
ftools} tasks. The event files of all the observations were filtered to include events
with {\sc flag}$=$0, and {\sc pattern}$\leq$12 ({\sc pattern}$\leq$4) and 200$\leq${\sc
pi}$\leq$12000 (150$\leq${\sc pi}$\leq$15000) for the MOS (pn) detectors.
The event files of the observations of SDSS~0210$-$0018, SDSS~0231$-$0728, and
PSS~0926$+$3055 were also filtered to remove periods of flaring activity, which were
apparent in the light curve of the entire full-frame window. Such flaring activity,
which originates in the Earth's magnetosphere, depends on several unpredictable
factors, and it can significantly reduce the good-time intervals of an
observation.\footnote{On average, $\sim$35\% of the data from \xmm\ observations
with exposure times greater than 10~ks are lost due to flaring (K. Kuntz, 2005,
private communication). For more details see the \xmm\ User Handbook at
http://xmm.vilspa.esa.es/external/xmm\_user\_support/ documentation/uhb/index.html \\}
The observations of SDSS~0210$-$0018, SDSS~0231$-$0728, and PSS~0926$+$3055
were shortened by 16~ks, 10~ks, and 0.6~ks, respectively, as a result of this
filtering. The event files of the PSS~0121$+$0347 and PSS~1326$+$0743
observations were not filtered in time, since the entire observation of the
first source was performed during a moderate background flare period with
count rates $\sim$3 times higher than nominal values, and the
full-frame count-rate light curve of the second source did not show any
flaring. The exposure times listed in Table~\ref{obs_log} reflect
the filtered data used in the analysis.

We searched for \hbox{X-ray} emission from an additional radio-quiet AGN,
SDSS~J021102.72$-$000910.3 at $z$=4.90 (Fan \et 1999), which lies close to the
edge of the field of view in the \xmm\ observation of SDSS~0210$-$0018. The former
source is not detected, and the 3$\sigma$ detection threshold is $\sim$120 counts.
Using {\sc pimms,}\footnote{Portable Interactive Multi-Mission Simulator at
http://heasarc.gsfc.nasa.gov/ Tools/w3pimms.html} this translates to an
upper limit on the unabsorbed observed-frame \hbox{0.5--2~keV} flux
of $\ltsim10^{-14}$~ergs~cm$^{-2}$~s$^{-1}$, which is consistent with the
3.1$^{+2.2}_{-1.4}\times$10$^{-15}$~ergs~cm$^{-2}$~s$^{-1}$ flux measured by
Vignali \et (2001).

\begin{figure}
\plotone{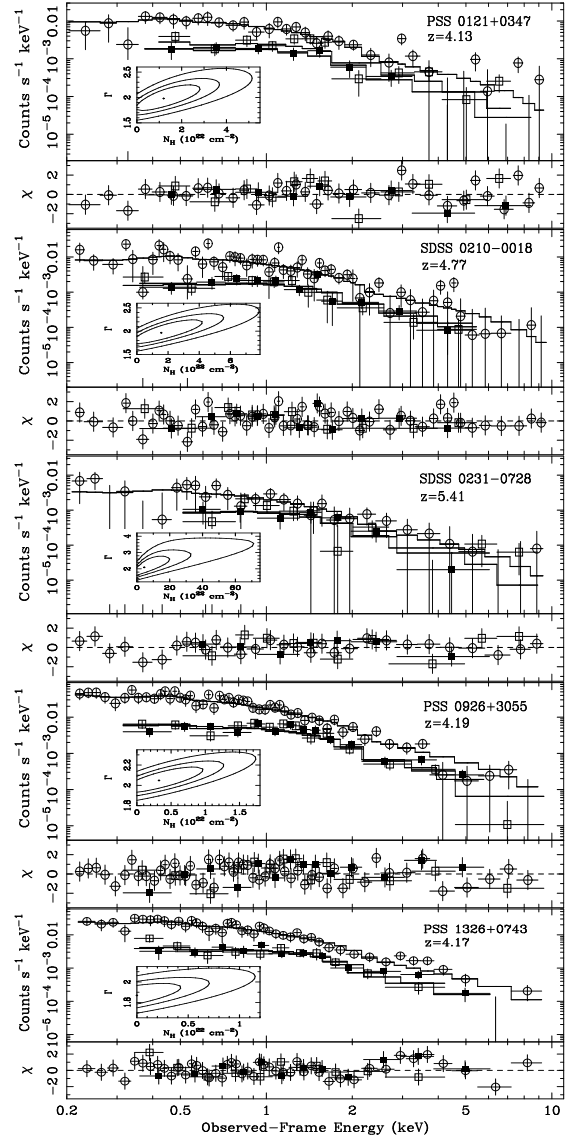}
\caption
{Data, best-fit spectra, and residuals for our new \xmm\ observations of five
\hbox{$z>4$} AGNs. Open circles, filled squares, and open squares represent the pn,
MOS1, and MOS2 data, respectively. Solid lines represent the best-fit model for
each spectrum, and the thick line marks the best-fit model for the pn data.
The $\chi$ residuals are in units of $\sigma$ with error bars of size one.
The inset in each panel shows 68, 90, and 99\% confidence contours for the
intrinsic absorption (\nh) and photon index ($\Gamma$).}
\label{spectra_fig}
\end{figure}
 
\begin{deluxetable*}{lcccccccc}
\tablecolumns{9}
\tablewidth{460pt}
\tablecaption{Fluxes, Luminosities, and \aox\ for the Sample of \hbox{$z>4$} AGNs
\label{opt_xray}}
\tablehead
{
\colhead{} &
\colhead{} &
\colhead{} &
\colhead{$f_{\nu}$[2500(1+$z$)~\AA]\tablenotemark{a}} &
\colhead{log $\nu L_{\nu}$(2500~\AA)} &
\colhead{Optical} &
\colhead{} &
\colhead{log $L_{{\rm 2-10~keV}}$\tablenotemark} &
\colhead{} \\
\colhead{AGN} &
\colhead{AB$_{{\rm 1450}(1+z)~\mbox{\rm \AA}}$} &
\colhead{$M_B$} &
\colhead{(mJy)} &
\colhead{(ergs s$^{-1}$)} &
\colhead{Ref.} &
\colhead{$F_{{\rm 0.5-2~keV}}$\tablenotemark{b}} &
\colhead{(ergs s$^{-1}$)} &
\colhead{\aox}
}
\startdata
PSS~0121$+$0347  & 18.5 & $-28.3$ &
$0.222\pm0.071$  & 46.94 & 1 &
$17.02^{+2.19}_{-2.28}$ & 45.51 &
$-1.65^{+0.03}_{-0.03}$ \\ \\
SDSS~0210$-$0018 & 19.3 & $-27.7$ &
$0.106\pm0.011$  & 46.71 & 2 &
$14.37^{+1.77}_{-1.76}$  & 45.27 &
$-1.54^{+0.03}_{-0.02}$ \\ \\
SDSS~0231$-$0728 & 19.3 & $-27.9$ &
$0.062\pm0.030$  & 46.56 & 3 &
$4.68^{+1.35}_{-1.29}$  & 45.21 &
$-1.62^{+0.06}_{-0.06}$ \\ \\
PSS~0926$+$3055  & 16.7 & $-30.1$ &
$1.167\pm0.042$  & 47.67 & 1 &
$39.04^{+2.57}_{-2.54}$ & 45.89 &
$-1.76^{+0.03}_{-0.01}$ \\ \\
PSS~1326$+$0743  & 17.2 & $-29.6$ &
$0.736\pm0.051$  & 47.46 & 1 &
$27.89^{+1.90}_{-2.52}$ & 45.73 &
$-1.76^{+0.03}_{-0.02}$ \\ \\
\hline \\
Q~0000$-$263& 17.5 & $-29.3$ &
$0.579\pm0.058$  & 47.35 & 4 &
$12.59^{+0.75}_{-0.75}$ &
45.72 &
$-1.70^{+0.03}_{-0.01}$ \\ \\
BR~0351$-$1034   & 18.7\tablenotemark{c} & $-28.2$ &
$0.185\pm0.019$  & 46.89 & 5 &
$11.69^{+2.22}_{-2.19}$ &
45.39 &
$-1.69^{+0.04}_{-0.04}$ \\ \\
SDSS~1030$+$0524 & 19.7 & $-27.8$ &
$0.076\pm0.008$  & 46.75 & 6 &
$3.09^{+0.53}_{-0.52}$  &
45.18 &
$-1.69^{+0.04}_{-0.04}$ \\ \\
BR~2237$-$0607   & 18.3\tablenotemark{c} & $-28.5$ &
$0.267\pm0.027$  & 47.08 & 5 &
$8.50^{+1.34}_{-1.27}$  &
45.32 &
$-1.74^{+0.04}_{-0.03}$ \\ \\
SDSS~1306$+$0356 & 19.6 & $-27.9$ &
$0.085\pm0.008$  & 46.76 & 6 &
$2.74^{+0.56}_{-0.53}$  &
45.07 &
$-1.75^{+0.05}_{-0.05}$ \\
\enddata
\tablenotetext{a}{Assuming an ultraviolet (UV) continuum of the form: $f_{\nu} \propto
\nu^{-0.79}$ (Fan \et 2001).}
\tablenotetext{b}{Galactic absorption-corrected flux in the observed \hbox{0.5--2~keV}
band in units of $10^{-15}$~ergs~cm$^{-2}$~s$^{-1}$.}
\tablenotetext{c}{Estimated using an empirical relation between the APM $R$ and
AB$_{{\rm 1450}(1+z)\mbox{\rm \tiny\AA}}$ magnitudes and redshift;
see \S~2 of Kaspi \et (2000).}
\tablerefs{(1) Vignali \et (2003a); (2) Vignali \et (2001);
(3) Vignali \et (2003b); (4) Schneider, Schmidt, \& Gunn (1989); (5) Storrie-Lombardi
\et (1996); (6) Brandt \et (2002)}
\end{deluxetable*}

\subsection{Spectral Analysis}
\label{analysis}

To extract \hbox{X-ray} spectra we used a source-extraction aperture radius of
30\arcsec\ in the images of all three European Photon Imaging Camera (EPIC) detectors
for all but one AGN, PSS~0121$+$0347. For this source we used an aperture radius of
20\arcsec\ for the source extraction, since a larger aperture was affected by the
(mainly hard X-ray) background flare, resulting in a hardening of the spectrum and
a lowering of the signal-to-noise ratio. For a nominal background level, increasing
the aperture radius from 20\arcsec\ to 30\arcsec\ should not have resulted in any
noticeable change in either photon index or normalization (Kirsch \et 2004).
Background regions were taken to be at least as large as the source regions;
these were annuli, when there were no neighboring sources, and circles otherwise.
The redistribution matrix files (RMFs; which include information on the detector
gain and energy resolution) and the ancillary response files (ARFs; which include
information on the effective area of the instrument, filter transmission,
detector window transmission and efficiency, and any additional energy dependent
effects) for the spectra were created with the {\sc sas} tasks {\sc rmfgen} and
{\sc arfgen}, respectively.\footnote{For more details see the NASA/GSFC calibration
memo CAL/GEN/92-002 at http://heasarc.gsfc.nasa.gov/docs/heasarc/caldb/docs/
memos/cal\_gen\_92\_002/cal\_gen\_92\_002.html} We note that the two reflection
grating spectrometer detectors on \xmm\ do not have sufficient counts to perform a
high-resolution spectral analysis for any of our sources.

The spectra were grouped with a minimum of 10--25 photons per bin using the task
{\sc grppha}. For each AGN we used {\sc xspec v11.3.0} (Arnaud 1996) to fit the
data with the following models: (i) a power-law model and a Galactic absorption
component (Dickey \& Lockman 1990), which was kept fixed during the fit, and
(ii) a model similar to the first with an added intrinsic (redshifted) neutral
absorption component. In all the fits we assumed solar abundances (Anders \& Grevesse
1989) and used the {\sc phabs} absorption model in {\sc xspec} with the
Balucinska-Church \& McCammon (1992) cross-sections. We fitted the data from each EPIC
detector alone, jointly fitted the data of the two MOS detectors, and jointly fitted
all three EPIC detectors with these models. In general, the five different fitting
results were consistent with each other within the uncertainties, and the joint fit of
all three EPIC detectors had the smallest uncertainty on the spectral parameters.
The best-fit results for the joint fit of the three EPIC detectors are summarized in
Table~\ref{best_fit} (first five sources). For all five sources the best-fit model is
a Galactic-absorbed power law; an $F$-test carried out on the data does not warrant
the addition of an intrinsic absorber in any of the spectra. The 90\% confidence upper
limits on intrinsic \nh\ appear in Table~\ref{best_fit}. The spectra of the five AGNs
and their best-fit models appear in Fig.~\ref{spectra_fig}. Also shown for each AGN
is a confidence-contour plot of the \hbox{$\Gamma$--\nh} parameter space. No strong
systematic residuals are seen for any of the fits. Optical fluxes and luminosities,
and \hbox{X-ray} fluxes and luminosities of the sources computed using the best-fit
spectral parameters of Table~\ref{best_fit}, appear in Table~\ref{opt_xray}
(first five sources).

\subsection{Serendipitous Sources}
\label{serendipitous}

Since the luminous AGNs under study are probably associated with massive potential
wells in the earliest collapsed large-scale structures, we searched for candidate
\hbox{$z>4$} AGNs in the \xmm\ images of three of our sources with SDSS imaging
(SDSS~0210$-$0018, SDSS~0231$-$0728, and PSS~1326+0743) to search for AGN clustering
on scales of \hbox{$\approx$0.1--5~Mpc} at \hbox{$z\sim$4--5} (e.g., see Djorgovski
\et 2003; Vignali \et 2003b; Hennawi \et 2005). We matched \hbox{X-ray} and SDSS
sources in those fields and utilized the Richards \et (2002) color criteria as well
as a pointlike morphology criterion to identify $\approx$3 \hbox{X-ray} emitting
$z\gtsim$4 AGN candidates for future optical spectroscopy. \\

\section{Archival Sample of \hbox{$z>4$} Radio-Quiet AGNs}
\label{expanded}

To improve the statistics on \hbox{X-ray} spectral properties, we expanded the
sample by adding moderate-to-high quality data of five radio-quiet \hbox{$z>4$} AGNs,
obtained from the \xmm\ and \chandra\ archives. A minimum of $\sim$200 ($\sim$100)
photons per source from \xmm\ (\chandra) observations was considered useful to our
analysis. The observation log of the four sources observed by \xmm\ appears in
Table~\ref{obs_log}, which also gives the references to the original papers presenting
the results of those observations. We processed the observation data files of these
sources following the analysis procedures described in \S~\ref{observations}. As in
the case of our original sample, we were only able to place upper limits upon the
intrinsic column densities in these sources. The best-fit model parameters for the
\hbox{X-ray} spectra of these sources are listed in Table~\ref{best_fit}.

The fifth observation, a 118~ks {\sl Chandra}/ACIS-S observation of
SDSS~J130608.26$+$035626.3 (hereafter SDSS~1306$+$0356) at $z$=5.99 in
2003 November 29--30, was re-analyzed following the same source and background
extraction parameters as those used in the original paper reporting that observation
(Schwartz \& Virani 2004) applying standard {\sc ciao\footnote{\chandra\ Interactive
Analysis of Observations. See http://asc.harvard.edu/ciao/} v3.2} routines to the data.
The events were grouped into bins that had a minimum of 15 counts per bin,
and the model fit to the data included a constant Galactic absorption
component (\nh=2.07$\times$10$^{20}$~cm$^{-2}$), a power-law component, and a
neutral \Ka\ line at rest-frame 6.4~keV with a width of $\sigma$=0.1~keV, since
there is a feature in the spectrum at that location hinting at the presence of
such a line (see Schwartz \& Virani 2004). The best-fit photon index is
$\Gamma$=1.82$^{+0.29}_{-0.31}$, the flux normalization at 1~keV is
(1.23$\pm$0.24)$\times10^{-6}$~keV~cm$^{-2}$~s$^{-1}$~keV$^{-1}$,
and the \Ka\ rest-frame equivalent width is 1.0$^{+1.1}_{-0.8}$~keV.

The optical fluxes and luminosities, and X-ray fluxes and luminosities of the five
sources, based upon the best-fit spectral parameters of Table~\ref{best_fit} and of
the single \chandra\ observation, are listed in Table~\ref{opt_xray}. Plots of the
spectra and best-fit models, which are generally consistent with our results, appear
in the original papers cited in Table~\ref{obs_log} and in Schwartz \& Virani (2004)
for the single \chandra\ observation.

\begin{figure}
\epsscale{0.9}
\plotone{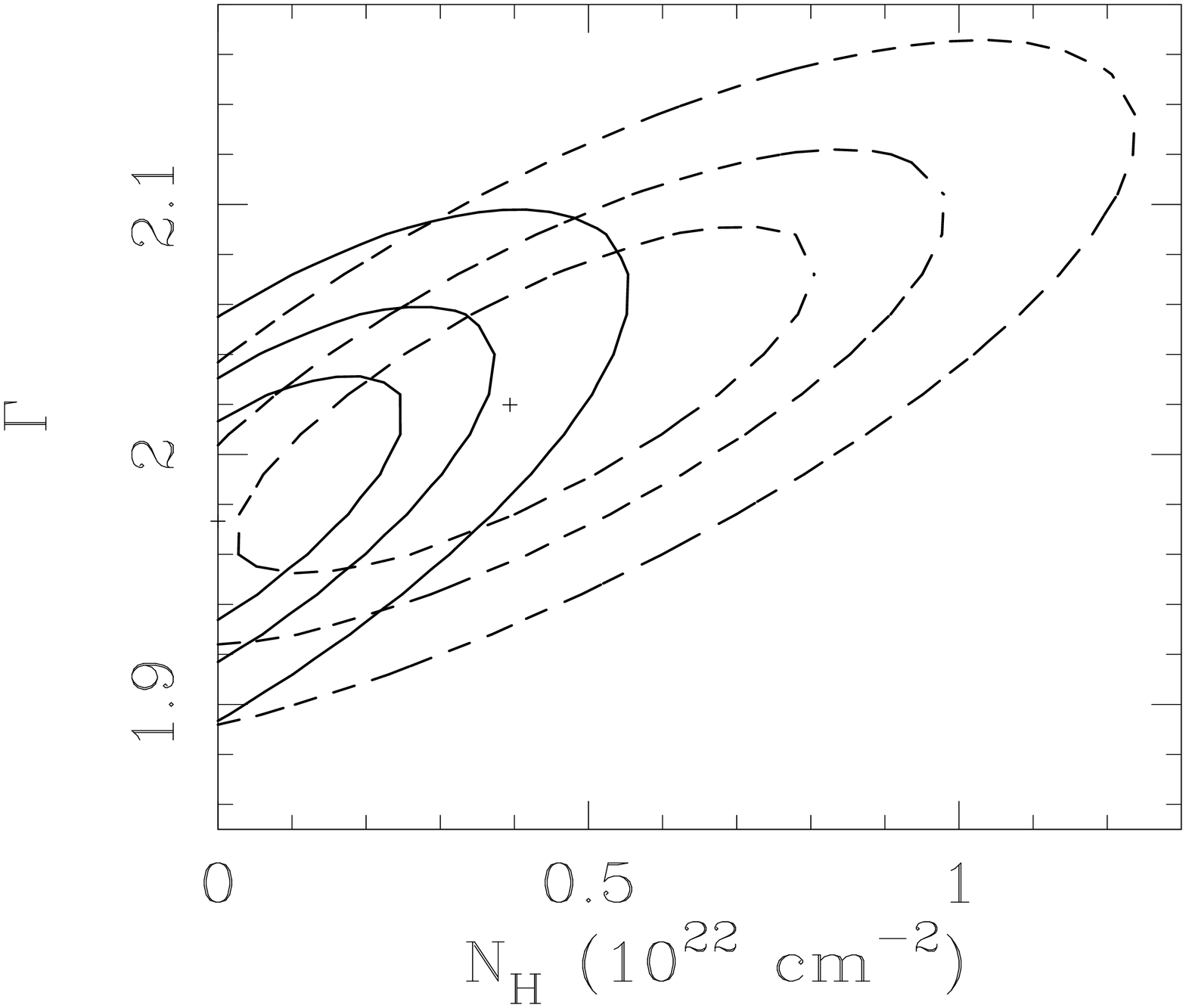}
\caption{68, 90, and 99\% confidence regions for the photon index vs.
intrinsic column density derived from joint spectral fitting of our sample
of eight radio-quiet AGNs. Solid (dashed) contours refer to the entire (common)
energy range of the AGNs (see \S~\ref{joint} for more details).}
\label{joint_fit}
\end{figure}

\section{Joint Spectral Fitting of \hbox{$z>4$} Radio-Quiet AGNs}
\label{joint}

To derive the best possible constraints on the average \hbox{X-ray} spectral
properties of radio-quiet \hbox{$z>4$} AGNs, we jointly fitted the spectra of the
expanded sample of eight radio-quiet AGNs (e.g., see \S~6 of Vignali \et 2005). We
combined the spectra from all the detectors of the observations of the eight
AGNs, 22 datasets in total, and used {\sc xspec} to fit their spectra jointly.
We fit two different models:
(i) a power-law that includes both Galactic and intrinsic absorption, and
(ii) a power-law with Galactic absorption including a Compton-reflection component,
using the {\sc pexrav} model in {\sc xspec} (Magdziarz \& Zdziarski 1995), and a
neutral narrow Gaussian \Ka\ line. The Compton-reflection component, also known as
a Compton-reflection `hump', is the spectral manifestation of hard \hbox{X-ray} photons
emitted in a corona of hot electrons and reflected off the relatively colder
accretion disk (e.g., see \S~3.5 of Reynolds \& Nowak 2003 and references therein).
The Compton-reflection hump feature is observed in the rest-frame \hbox{$\sim$7--60~keV}
energy range, peaking at rest-frame $\sim$30~keV.
The joint-fitting process was carried out twice for both models;
first, to include the entire energy range of all the datasets, and
a second time to include only the common rest-frame energy range of the
eight sources: 1.5$\ltsim E_{\rm rest}\ltsim$51~keV.
The total number of photons used in the joint-fitting
process is $\sim$7000 ($\sim$6400 for the common energy range), and
it is larger by an order of magnitude than the number of counts used in the
joint fitting of 48 \hbox{$z>4$} radio-quiet AGNs by Vignali \et (2005).

\begin{deluxetable*}{lcccccc}
\tablecolumns{7}
\tablewidth{460pt}
\tablecaption{Best-Fit Parameters from Joint Fitting of Eight Radio-Quiet \hbox{$z>4$}
AGNs
\label{joint_fit_par}}
\tablehead
{
\colhead{Spectral Model} &
\colhead{Energy} &
\colhead{\nh} &
\colhead{} &
\colhead{EW(\Ka)\tablenotemark{b}} &
\colhead{} &
\colhead{} \\
\colhead{Galactic-Absorbed Power Law and} &
\colhead{Range\tablenotemark{a}} &
\colhead{(10$^{22}$~cm$^{-2}$)} &
\colhead{$\Gamma$} &
\colhead{(eV)} &
\colhead{$R$\tablenotemark{c}} &
\colhead{$\chi^{2}$(DOF)}
}
\startdata
Intrinsic Absorption           & E &
$\le0.28$ &
$1.97^{+0.06}_{-0.04}$ & \nodata   &
\nodata                & 308.3 (360) \\ \\
Intrinsic Absorption           & C &
$\le0.84$ &
$2.02^{+0.04}_{-0.06}$ & \nodata   &
\nodata                & 291.3 (338) \\ \\
Compton-Reflection and neutral \Ka\ line   & E &
\nodata   &
$1.98^{+0.10}_{-0.09}$ & $\le$189   &
$\le1.22$              & 306.0 (359) \\ \\
Compton-Reflection and neutral \Ka\ line   & C &
\nodata   &
$1.95^{+0.10}_{-0.08}$ & $\le$191   &
$\le1.03$              & 291.2 (337) \\
\enddata
\tablenotetext{a}{E - entire energy range of all the spectra. C - the common
rest-frame energy range among all sources (\hbox{1.5--51~keV}).}
\tablenotetext{b}{Rest-frame equivalent width calculated with a mean
redshift of 4.8.}
\tablenotetext{c}{Compton hump relative reflection parameter
(see Magdziarz \& Zdziarski 1995).}
\end{deluxetable*}

The fitting process with the first model required that all datasets be
fitted with a single power-law and a single intrinsic absorption column. Therefore
the photon index and intrinsic absorption column were each tied to a single
value across all data sets, and these two parameters were allowed to vary
jointly (see the discussion of intrinsic dispersion in \S~\ref{index} as partial
justification of this approach). The power-law normalizations of all data sets
were allowed to vary freely. For the second model we used a single power-law and
a single relative-reflection component ($R$, where $R$ is defined as $\Omega/2\pi$
and $\Omega$ is the solid angle subtended by the reflecting medium) of the Compton
hump for all datasets, and allowed them to vary jointly. We fixed the rest-frame
energy and width ($\sigma$) of the neutral \Ka\ line to 6.4 and 0.1~keV, respectively,
and tied all the power-law and Gaussian normalizations to two different values,
allowing them to vary jointly; each dataset was assigned a constant scaling factor
to account for the different flux levels of each AGN and each detector.
The best-fit values and upper limits for the parameters used in the joint fits are
given in Table~\ref{joint_fit_par}. Fig.~\ref{joint_fit} is a contour plot of
$\Gamma$ versus \nh\ as a result of the joint-fitting process using the first model.
The noticeable difference between the confidence contours in the entire and common
energy ranges stems from the fact that the two AGNs at $z\simeq$6 increase the
low-energy threshold in the common energy range, and as a consequence many low-energy
photons from the \hbox{X-ray} bright AGNs at $z\simeq$4.2 are ignored in the fit. This
causes the upper limit on the intrinsic absorption in the common energy range to rise
significantly relative to the upper limit when using the entire energy range.

We repeated the joint-fitting process with the first model and fitted the
four most optically luminous radio-quiet sources separately from the four least
optically luminous radio-quiet sources both for the entire and common energy ranges
of the two sub-samples. Both $\Gamma$ and \nh\ in the two luminosity groups are
consistent with one another and with the values we obtain by fitting the entire
radio-quiet sample. For the fitting using the entire energy range, we measure
$\Gamma$=1.99$\pm$0.06 for the most-luminous sources and
$\Gamma$=1.95$^{+0.23}_{-0.20}$ for the least-luminous sources.

\begin{figure*}
\plotone{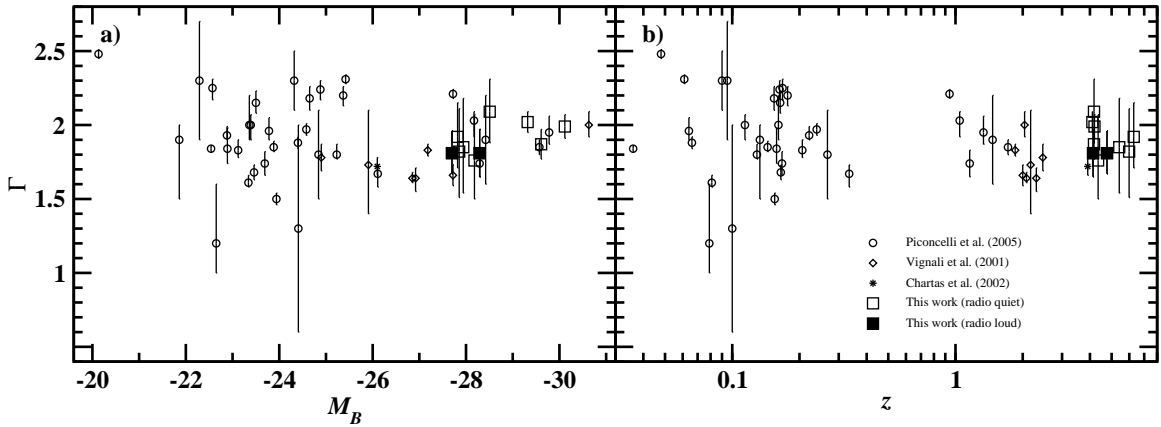}
\caption{The \hbox{X-ray} photon index ($\Gamma$), versus ({\it a}) absolute $B$
magnitude and ({\it b}) redshift, adapted from Vignali \et (2005).
The eight (two) radio-quiet (moderately radio-loud) AGNs presented in this work are
marked with open (filled) squares. Note that the photon index does not
show any clear dependence on either luminosity or redshift.}
\label{Gamma_MB_z}
\end{figure*}

\section{Results and Discussion}
\label{discussion}
\subsection{X-ray Photon Index}
\label{index}

The photon indices and their uncertainties for the full sample of ten AGNs, in the
\hbox{$\sim$1--50~keV} rest-frame energy range (the range where most photons are
detected), are listed in Table~\ref{best_fit}. A statistical summary of our expanded
sample's photon-index distribution appears in Table~\ref{Gamma}. Using the maximum
likelihood method of Maccacaro \et (1988) we find no significant intrinsic dispersion
in $\Gamma$; this is expected since the individual uncertainties on $\Gamma$ are
larger than the spread in individual $\Gamma$ values. In Fig.~\ref{Gamma_MB_z} we plot
$\Gamma$ versus absolute $B$ magnitude and redshift for our expanded sample, together
with previous measurements of the photon index in other AGN samples (see \S~6 of
Vignali \et 2005 and references therein). There is no obvious systematic difference
between the photon indices of our radio-quiet sources and those of luminous AGNs
observed at lower redshifts (e.g., the Piconcelli \et 2005 sample of nearby AGNs).
Using a Spearman rank correlation method we find no significant correlations between
$\Gamma$ in radio-quiet sources and either luminosity or redshift as also found at lower
redshift ($z\ltsim$0.4) by, e.g., Porquet \et (2004). Applying a linear regression
algorithm that accounts for intrinsic scatter and measurement errors (Akritas \&
Bershady 1996), we obtain $|\partial\Gamma/\partial z|<0.04$ for the 51 radio-quiet
AGNs plotted in Fig.~\ref{Gamma_MB_z}. From the joint fitting of eight radio-quiet AGN
spectra using the absorbed power-law model we find $\Gamma=1.97^{+0.06}_{-0.04}$. This
is consistent with both the unweighted mean of the individual $\Gamma$ values and the
counts-weighted mean $\Gamma$ for the radio-quiet sample.

The mean photon indices we find for \hbox{$z>4$} radio-quiet AGNs are also all
consistent with those calculated by Vignali \et (2005). All these photon-index values
confirm and strengthen the Vignali \et (2005) result that the \hbox{X-ray} photon index
does not depend on optical luminosity or redshift. In particular, our results do not
support either of the reports for a significantly different photon index at low versus
high redshift (Bechtold \et 2003; Grupe \et 2005). Bechtold \et (2003) used a
heterogeneous sample of low-redshift {\sl ROSAT} sources and high-redshift sources
detected with \chandra. The mix of different datasets may have produced a bias
which led Bechtold \et (2003) to report a flattening of $\Gamma$ at higher
redshift (e.g., due to spectral curvature at low energies and mission-to-mission
cross-calibration uncertainties). In addition, the counts-weighted mean $\Gamma$ of
their \hbox{$z>4$} \chandra\ sources is 1.50$\pm$0.15 (see their Table~2). This is
significantly lower than the average $\Gamma$ of 1.93$^{+0.10}_{-0.09}$ for 48
radio-quiet AGNs, including the Bechtold \et (2003) sources, obtained by Vignali \et
(2005). Comparing the photon indices of our four sources which overlap with those
of Grupe \et (2005) shows that in three cases the values are in good agreement.
However, for SDSS~1030$+$0524 the $\Gamma$=2.65$\pm$0.34 found by Grupe \et (2005)
is significantly steeper, and is inconsistent both with the value we obtain,
$\Gamma$=1.92$^{+0.22}_{-0.21}$, and that obtained by Farrah \et (2004),
$\Gamma$=2.12$\pm$0.11. The photon index we obtain for SDSS~1030$+$0524, while
slightly harder, is compatible within the errors with the one found by Farrah \et
(2004). By inspection of Tables~1~and~4 of Grupe \et (2005) it is also apparent that
the relatively steep mean $\Gamma$ they report for their radio-quiet sources,
2.23$\pm$0.48, is an unweighted mean dominated by sources with limited
photon statistics ($\approx$50 counts), which in most cases result in best-fit
$\Gamma$ exceeding a value of 2.2.

The insignificant dispersion in individual $\Gamma$ values for our \hbox{$z>4$} AGN
sample may partially be driven by a selection bias, since we specifically
targeted some of the most \hbox{X-ray} luminous AGNs known. However, observing down the
AGN luminosity function at \hbox{$z>4$} is challenging with
existing \hbox{X-ray} observatories. In fact, the faintest \hbox{X-ray} sources at those
redshifts with enough counts to allow meaningful measurement of the photon
index were detected in the 2~Ms \chandra\ Deep Field North
survey (Vignali \et 2002; see Table~1 of Brandt \et 2005 for a list of X-ray
faint AGNs at \hbox{$z>4$}). The best studied source in that survey,
CXOHDFN~J123647.9$+$620941 at $z$=5.19 and with
$L_{2-10~\rm keV}$=10$^{44.0}$~ergs~s$^{-1}$, has $\Gamma$=1.8$\pm$0.3, which is
consistent with the photon indices we find for our sources (Vignali \et 2002).
The photon indices of our least luminous sources do not appear to be significantly
different from those of the more luminous ones both individually and on average by
splitting the joint spectral fit into two luminosity groups (see \S~\ref{joint}).
Inspection of Table~4 of Vignali \et (2005) also does not show a strong dependence of
$\Gamma$ on optical luminosity. In spite of being extremely luminous, our \hbox{$z>4$}
AGNs are otherwise representative of the AGN population by having `normal' rest-frame
UV properties (i.e., no broad absorption lines, typical emission-line
properties and rest-frame UV continua; Vignali \et 2001, 2003a, b). Recent
multiwavelength observations of \hbox{$z>4$} AGNs have found that their
overall spectral energy distributions (SEDs) are not significantly different from
those of lower redshift sources implying no SED evolution (e.g., see Carilli \et
2001 and Petric \et 2003 for radio observations; Pentericci \et 2003 and Vanden Berk
\et 2001 for \hbox{UV--optical} observations). We argue that the broad-band X-ray
spectral properties of \hbox{$z>4$} AGNs are also not significantly different than
those of lower redshift sources.

\begin{deluxetable}{lcc}
\tablecolumns{3}
\tablecaption{Photon Index Statistics of \hbox{$z>4$} AGNs
\label{Gamma}}
\tablehead
{
\colhead{} &
\multicolumn{2}{c}{Photon Index ($\Gamma$)} \\
\colhead{Statistic} &
\colhead{Entire Sample} &
\colhead{Radio-Quiet Sources}
}
\startdata
Unweighted Mean                   & $1.89$ & $1.92$ \\
Unweighted Standard Deviation     & $0.11$ & $0.11$ \\
Unweighted Error of the Mean      & $0.03$ & $0.04$ \\
Counts-Weighted Mean              & $1.93$ & $1.95$ \\
Median                            & $1.86$ & $1.90$
\enddata
\end{deluxetable}

\subsection{Intrinsic Absorption}
\label{sec_nh}

In all the individual fits of the \hbox{$z>4$} AGNs we obtained an intrinsic \nh\
consistent with zero with 90\% confidence. Table~\ref{best_fit} lists the upper limits
on the intrinsic absorption; they are all in the range
$\approx$10$^{22}$--10$^{23}$~cm$^{-2}$. The joint-fitting procedure yielded an upper
limit on the common intrinsic \hbox{\nh~of~$\sim$(3--8)$\times10^{21}$~cm$^{-2}$}
(Table~\ref{joint_fit_par} and Fig.~\ref{joint_fit}). These constraints on intrinsic
absorption are the tightest presented to date for \hbox{$z>4$} AGNs, and our
individual-fitting results are less prone to bias than the joint-fitting results that
have been predominantly utilized previously.\footnote{Since photoelectric absorption
by neutral gas is an extremely strong spectral effect, a column density derived via
joint fitting of several sources may not always be a good estimate of their `typical'
column density. The presence of even one source with significant low-energy flux can
bias the joint-fitting column density downward, since such flux is highly inconsistent
with a strong photoelectric-absorption cutoff at low energy.} Our results show that
radio-quiet AGNs at \hbox{$z>4$} do not appear to have higher absorption columns than
their lower redshift counterparts. The metallicity in the inner regions of
high-redshift AGNs may be higher, on average, than in their lower-redshift counterparts
mainly due to their higher luminosities (e.g., Dietrich \et 2003). Since \hbox{X-ray}
photoelectric absorption is predominantly due to metals and not hydrogen, the inferred
intrinsic neutral hydrogen column density depends upon the assumed metallicity.
Therefore, \nh\ values are inversely proportional to the assumed metallicity, and the
quoted upper limits on the best-fit intrinsic columns given in
Table~\ref{joint_fit_par} are conservative values (as solar abundances were assumed).

Despite the considerable large-scale evolution the Universe has experienced
since $z\approx$6, and in particular the strong evolution in the AGN population
(e.g., Fan \et 2003; Croom \et 2004; Barger \et 2005; Brandt \& Hasinger 2005),
we do not detect significant changes in the central regions of luminous radio-quiet
AGNs. Since the host galaxies of AGNs at high redshifts are probably still in
the formation process, one might expect large amounts of gas and dust in the
interstellar medium of these hosts that could increase the amount of line-of-sight
absorption. Signatures of large amounts of interstellar molecular gas and dust
have been recently detected in some of the most distant AGNs
(e.g., Carilli \et 2001; Bertoldi \et 2003; Walter \et 2003; Maiolino \et 2004).
In addition, the fact that radio-loud AGNs show signs for increased intrinsic
absorption with redshift (see \S~\ref{introduction}), makes it of interest to check
if there is a similar trend for the radio-quiet population. Our results do not
show significant intrinsic absorption in any of the \hbox{$z>4$} radio-quiet AGNs, which
is in broad agreement with, e.g., the Laor \et (1997) results for local radio-quiet
AGNs. We caution that our AGN sample is far from being complete or representative
of the entire AGN population. We have specifically considered only optically
and \hbox{X-ray} bright, radio-quiet AGNs with no broad absorption lines. Steffen
\et (2003) have found that the fraction of type 1 (i.e., unobscured) AGNs
increases with \hbox{X-ray} luminosity, suggesting luminosity-dependent absorption.
It is therefore possible that the relatively lower intrinsic absorption in our
sources is a consequence of them being among the most luminous AGNs known.

\subsection{\Ka\ Line and Compton Reflection}
\label{sec_ka}

The joint-fitting process placed respectable constraints on the strength of
a putative neutral narrow \Ka\ line; a 90\% confidence upper limit on the
rest-frame equivalent width is $\sim$190~eV. We also constrained the relative
reflection component of a Compton-reflection hump (see Table~\ref{joint_fit_par}).
These are the best constraints to date on these properties of \hbox{$z>4$} AGNs. Due to
the high mean \hbox{X-ray} luminosity ($\langle {\rm Log}~L_{{\rm 2-10~keV}}
({\rm ergs~s}^{-1}) \rangle \simeq$45.5) of our sample, both the \Ka\ line and
Compton hump are expected to be relatively weak. Using the recent Page \et
(2004a) anticorrelation between EW(\Ka) and $L_{{\rm 2-10~keV}}$
(the so-called \Ka\ ``Baldwin Relation''), we find an empirically expected
range of EW(\Ka)$\sim$25--80~eV for our sources, several times lower than our upper
limit (but see also Jimenez-Bail{\' o}n \et 2005). Given our results, the
EW(\Ka)$\approx$1~keV for SDSS~1306$+$0356 (see \S~\ref{expanded}) appears
remarkable and should be investigated with better data. The constraints we obtain on
the Compton-reflection component, $R\ltsim$1.2, are consistent with those obtained for
some \hbox{$z\simeq$0--1} AGNs (e.g., Page \et 2004b), but are significantly lower
than the value $R$=2.87$\pm$0.96 found in PG~1247+267 at $z$=2.04, the most luminous
source in which such a component was significantly detected (Page \et 2004b).

\begin{figure}
\epsscale{0.9}
\plotone{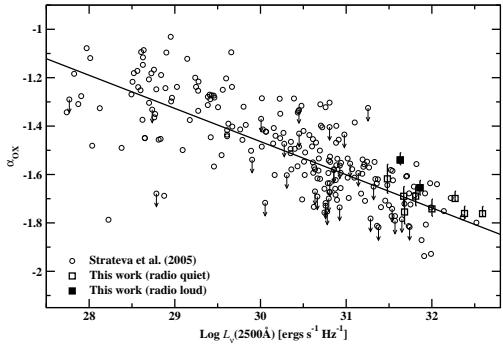}
\caption{The optical-to-X-ray SED parameter, \aox, versus luminosity density at
2500~\AA. The ten AGNs studied in this work ({\it squares}) follow the Strateva
\et (2005) \aox--$L_{\nu}$(2500~\AA) relation ({\it solid line}),
\aox$=-0.136\times L_{\nu}$(2500~\AA)$+2.616$. The two moderately radio-loud sources
presented in this work, PSS~0121$+$0347 and SDSS~0210$-$0018, are the only 
such sources on this diagram and are marked with filled squares. The reason why
most of our sources lie above the best-fit line may be attributed to the fact that
we specifically targeted \hbox{X-ray} bright sources.}
\label{aox}
\end{figure}

\subsection{Optical-to-X-ray SED}
\label{sec_aox}

For each of the ten AGNs we computed the optical-to-X-ray SED parameter, \aox,
defined~as:

\begin{equation}
\alpha_{\rm ox}=\frac{\log(f_{\rm 2~keV}/f_{2500\mbox{\rm~\scriptsize\AA}})}
{\log(\nu_{\rm 2~keV}/\nu_{2500\mbox{\rm~\scriptsize\AA}})}
\end{equation}
where $f_{\rm 2~keV}$ and $f_{2500\mbox{\rm~\scriptsize\AA}}$ are the flux densities at
rest-frame 2~keV and 2500~\AA, respectively (e.g., Tananbaum \et 1979).

The \aox\ values and their 1$\sigma$ errors appear in Table~\ref{opt_xray}. The errors
on \aox\ were computed following the ``numerical method'' described in \S~1.7.3 of
Lyons (1991), taking into account the uncertainties on the \hbox{X-ray} power-law
normalizations and photon indices, and the effects of possible changes in the
\hbox{UV--optical} slope (from $\alpha$=$-$0.5 to $-$0.79). In Fig.~\ref{aox} we plot
\aox\ for our sources against their optical luminosities, and include a sample of 228
radio-quiet AGNs (mainly from the SDSS) from Strateva \et (2005) in the diagram. Also
plotted in Fig.~\ref{aox} is the best-fit line of the Strateva \et (2005)
\hbox{\aox--$L_{\nu}$(2500~\AA)} relation (Eq.~6 of Strateva \et 2005).
The \aox\ values of the ten AGNs presented in this work, which represent
extreme luminosities and redshifts, generally follow that relation,
although most of them lie above the best-fit line. This may be due to a selection
bias since we have specifically targeted \hbox{X-ray} bright sources.

\begin{figure}
\epsscale{1.2}
\plotone{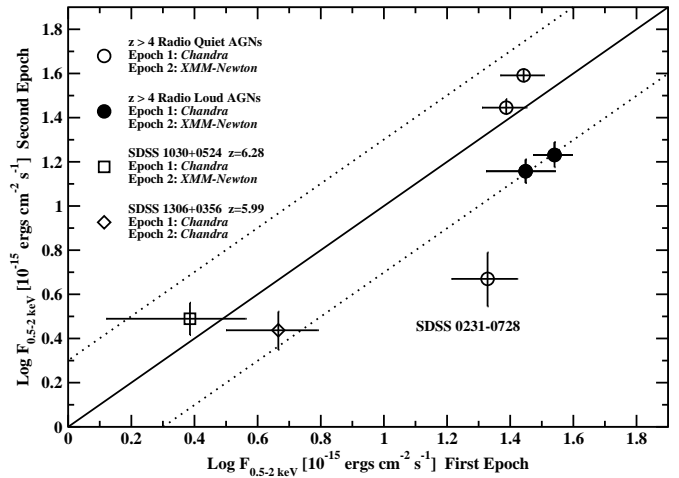}
\caption
{Two-epoch Galactic-absorption corrected \hbox{0.5--2~keV} fluxes for seven
of the \hbox{$z>4$} AGNs in our sample. The solid line marks the 1:1 flux
ratio, and the two dotted lines mark 1:2, and 2:1 flux ratios, to guide
the eye. SDSS~0231$-$0728 clearly varied by more than a factor of two between
the two epochs. The second and third most variable sources, PSS~0121$+$0347
and SDSS~0210$-$0018, are moderately radio loud and are marked with filled circles.}
\label{z4_var}
\end{figure}

\begin{deluxetable*}{lcccccc}
\tablecolumns{7}
\tablewidth{460pt}
\tablecaption{Two-Epoch \hbox{X-ray} Fluxes Obtained by \xmm\ and \chandra
\label{two_epoch}}
\tablehead
{
\colhead{} &
\multicolumn{2}{c}{First Epoch} &
\multicolumn{2}{c}{Second Epoch} &
\colhead{} &
\colhead{} \\
\colhead{} &
\colhead{$F_{{\rm 0.5-2~keV}}$\tablenotemark{a}} &
\colhead{} &
\colhead{$F_{{\rm 0.5-2~keV}}$\tablenotemark{a}} &
\colhead{} &
\colhead{$\Delta t$\tablenotemark{b}} \\
\colhead{AGN} &
\colhead{(10$^{-15}$~ergs~cm$^{-2}$~s$^{-1}$)} &
\colhead{Ref.} &
\colhead{(10$^{-15}$~ergs~cm$^{-2}$~s$^{-1}$)} &
\colhead{Ref.} &
\colhead{(days)} &
\colhead{$\chi^2$\tablenotemark{c}}
}
\startdata
PSS~0121$+$0347  & $34.7^{+5.4}_{-4.4}$   & 1 & $17.02^{+2.19}_{-2.28}$ &
2 & 135 & 20.40 \\ \\
SDSS~0210$-$0018 & $28.1^{+8.1}_{-6.0}$   & 3 & $14.37^{+1.77}_{-1.76}$ &
2 & 258  & 16.35 \\ \\ 
SDSS~0231$-$0728 & $21.3^{+5.6}_{-4.6}$   & 4 & $4.68^{+1.35}_{-1.29}$  &
2 & 73  & 41.42 \\ \\ 
PSS~0926$+$3055  & $27.7^{+4.7}_{-4.1}$   & 1 & $39.04^{+2.57}_{-2.54}$ &
2 & 186 & 6.43 \\ \\
PSS~1326$+$0743  & $24.4^{+4.3}_{-3.6}$   & 1 & $27.89^{+1.90}_{-2.52}$ &
2 & 139 & 0.64 \\ \\
SDSS~1030$+$0524 & $2.43^{+1.48}_{-0.99}$ & 5 & $3.09^{+0.53}_{-0.52}$  & 
6 & 68  & 0.45 \\ \\
SDSS~1306$+$0356 & $4.63^{+1.77}_{-1.33}$ & 5 & $2.74^{+0.56}_{-0.53}$  &
7 & 96  & 3.35 \\
\enddata
\tablenotetext{a}{Galactic absorption-corrected flux in the observed 0.5--2
keV band. Errors represent 90\% confidence limits on the fluxes.}
\tablenotetext{b}{Rest-frame time difference between the two epochs.}
\tablenotetext{c}{Variability statistic; see \S~\ref{variability}.}
\tablerefs{(1) Vignali \et (2003a); (2) This work; (3) Vignali \et (2001);
(4) Vignali \et (2003b); (5) Brandt \et (2002); (6) Farrah \et (2004);
(7) Schwartz \& Virani (2004)}
\end{deluxetable*}

\subsection{X-ray Variability}
\label{variability}

Using the event files of our \xmm\ observations, we searched for rapid
($\sim$1~hr timescale in the rest frame) variability of the AGNs by applying
Kolmogorov-Smirnov tests to the lists of photon arrival times. No
significant variations were detected, and the fractional variability of
the count-rate light curves was smaller than 15\% in all cases.
To test if the \hbox{X-ray} flux exhibits long-term (months-to-years) variations
in our sample, we list in Table~\ref{two_epoch} pairs of two-epoch \hbox{0.5--2~keV}
fluxes for seven \hbox{$z>4$} sources which were observed with \chandra\ in the first
epoch, and with either \xmm\ or \chandra\ in the second epoch.
The relative uncertainties between flux measurements obtained with \xmm\
versus those obtained with \chandra\ are considered to be the smallest
among different \hbox{X-ray} observatories, and are normally $\ltsim$10\%
(Snowden 2002; S. Snowden, 2005, private communication). Table~\ref{two_epoch} also
lists the rest-frame temporal separation between the two observations, and a $\chi^2$
statistic to quantify the significance of variation between the two epochs. The null
hypothesis is that the flux in each epoch is equal to the mean flux of the two epochs.
A larger $\chi^2$ implies higher significance of variability, and values larger than
2.71 (6.63) imply that the two fluxes are inconsistent within their uncertainties,
i.e., there is significant variability at a 90\% (99\%) confidence level (e.g., Avni
1976). Of the seven AGNs studied only two sources, PSS~1326$+$0743 and
SDSS~1030$+$0524, do not show a significant (i.e., at $\geq$90\% confidence) long-term
variation. Fig.~\ref{z4_var} plots the fluxes of the second epoch against those of the
first, as given in Table~\ref{two_epoch}.

While most AGNs varied by no more than a factor of $\approx$2 between the two epochs
one source, SDSS~0231$-$0728, faded by a factor of $\sim$4 between the first (\chandra)
observation and the second (\xmm) one. This flux change occurred over a rest-frame
period of 73~d. This is the largest change in \hbox{X-ray} flux observed in a
\hbox{$z>4$} radio-quiet AGN (e.g., Paolillo \et 2004). Given the optical flux of the
source, and using the Vignali \et (2003b) relation between optical and \hbox{X-ray}
fluxes, it is likely that this source was caught in an \hbox{X-ray} high state in the
first epoch (Vignali \et 2003b), since its \hbox{X-ray} flux in the second epoch
(this work) agrees with the value predicted from its optical flux (assuming the optical
flux is nearly constant). Vignali \et (2003b) also noted that SDSS~0231$-$0728
was \hbox{X-ray} brighter than expected (see their Fig.~5). The spectral slope of the
source also shows a possible indication of flattening from $\Gamma$=2.8$^{+1.10}_{-0.95}$
to $\Gamma$=1.85$^{+0.33}_{-0.31}$ between the two epochs, but the significance is only
$\sim$1$\sigma$ due to the limited number of counts ($\sim$25) in the first \chandra\
snapshot observation. This is a tentative indication for a transition from a soft/high
state to a hard/low state in this source, as has been seen for a few local AGNs
(e.g., Guainazzi \et 1998; Maccarone, Gallo, \& Fender 2003).

\section{Summary and Future Prospects}
\label{summary}

Moderate-to-high quality \hbox{X-ray} spectra of ten luminous AGNs at \hbox{$z>4$},
including five new \xmm\ observations, have been analyzed to extract \hbox{X-ray}
spectral properties and search for flux variations. We found that the \hbox{X-ray}
power-law photon index does not depend on either luminosity or redshift, broadly
consistent with the lack of multiwavelength AGN SED evolution, and it has no detectable
intrinsic dispersion at \hbox{$z>4$}. Upper limits placed on the intrinsic absorption
in eight radio-quiet \hbox{$z>4$} AGNs show that, on average, the \hbox{X-ray} light
of luminous radio-quiet AGNs at \hbox{$z>4$} is not more absorbed than in nearby AGNs.
All this suggests that the \hbox{X-ray} production mechanism and the central
environment in radio-quiet AGNs do not depend on luminosity and have not evolved
over cosmic time. We also place constraints on the strength of a putative narrow \Ka\
line, and constrain the relative strength of the Compton-reflection hump at high
rest-frame \hbox{X-ray} energies. Tighter constraints on those X-ray properties of
luminous AGNs at \hbox{$z>4$}, with existing \hbox{X-ray} observatories, can only be
achieved by lengthy (several 100~ks) observations of the most luminous sources.

A search for short-timescale ($\sim$1~hr) X-ray variability found no significant
variations, but on longer timescales (months-to-years) we have found significant
variations in five of our sources, noting one extreme radio-quiet source in which the
\hbox{X-ray} flux dropped by a factor of $\sim$4 over a 73~d rest-frame period; this
is the most extreme \hbox{X-ray} variation ever observed for a radio-quiet AGN at
\hbox{$z>4$}. X-ray monitoring of SDSS~0231$-$0728 and other AGNs at high redshift will
enable us to link the variability properties of high-redshift AGNs with some of their
fundamental properties, such as BH mass, luminosity, and accretion rate, as is done at
lower redshift (e.g., O'Neill \et 2005). Two recent studies have reported that the
\hbox{X-ray} flux variations of high-redshift AGNs are larger than those of their
nearby counterparts (Manners \et 2002; Paolillo \et 2004). The data presented in this
paper do not allow us to test this claim. Multiple-epoch \hbox{X-ray} observations of
a large sample of high-redshift radio-quiet AGNs are required to investigate their
variability properties, and compare them with those of the local AGN population.

\acknowledgments

This work is based on observations obtained with \xmm, an ESA science mission with
instruments and contributions directly funded by ESA Member States and the USA (NASA).
We are grateful to Bret Lehmer and Divas Sanwal for help with the \xmm\ data reduction.
We also thank Aaron Steffen, Iskra Strateva, and Dan Vanden Berk for useful discussions.
An anonymous referee is gratefully acknowledged for helping to improve the presentation
of this paper. We gratefully acknowledge the financial support of NASA grant
NNG04GF57G (OS, WNB, DPS), NASA LTSA grant NAG5-13035 (WNB, DPS), and NSF grants
AST-0307582 (DPS), AST-0307409 (MAS), and AST-0307384 (XF). XF acknowledges
support from an Alfred P. Sloan Research Fellowship, and a David and Lucile
Packard Fellow in Science and Engineering. CV acknowledges support from
MIUR cofin grant 03-02023 and INAF/PRIN grant 270/2003.

\end{document}